\documentclass[preprint,verbatim]{emulateapj}
\usepackage{graphicx}

\begin{document}

\title{Revised Pulsar Spindown}

\author{I. Contopoulos\altaffilmark{1} and A. Spitkovsky \altaffilmark{2,3}}

\altaffiltext{1}{Research Center for Astronomy, Academy of Athens, Greece}
\altaffiltext{2}{ Kavli Institute for Particle Astrophysics and Cosmology, Stanford University, P.O. Box 20450, Stanford, Ca 94309}
\altaffiltext{3}{Chandra Fellow}

\begin{abstract}
We address the issue of electromagnetic pulsar spindown by combining our
experience from the two limiting idealized cases which have been studied in
great extent in the past: that of an aligned rotator where ideal MHD conditions
apply, and that of a misaligned rotator in vacuum.  We construct a spindown
formula that takes into account the misalignment of the magnetic and rotation
axes, and the magnetospheric particle acceleration gaps.  We show that near the
death line aligned rotators spin down much slower than orthogonal ones.  In
order to test this approach, we use a simple Monte Carlo method to simulate the
evolution of pulsars and find a good fit to the observed pulsar distribution in
the $P-\dot{P}$ diagram without invoking magnetic field decay. Our model may
also account for individual pulsars spinning down with braking index $n<3$, by
allowing the corotating part of the magnetosphere to end inside the light
cylinder. We discuss the role of magnetic reconnection in determining the
pulsar braking index.  We show, however, that $n\sim 3$ remains a good
approximation for the pulsar population as a whole. Moreover, we predict that
pulsars near the death line have braking index values $n> 3$, and that the
older pulsar population has preferentially smaller magnetic inclination
angles. We discuss possible signatures of such alignment in the existing pulsar
data.
\end{abstract}

\maketitle

\section{Introduction}

The current canonical pulsar paradigm is that of a magnetized
rotating neutron star (see Mestel~1999 for a review). 
However, we feel that certain
fundamental aspects of the paradigm still remain unclear.

One aspect of the paradigm which we hope to elucidate in the present work has
to do with the way the neutron star spins down. 
A spinning down neutron star with mass $M_*$, radius $r_*$,
and angular velocity $\Omega$ loses rotational kinetic energy at a rate
\begin{equation}
L= \frac{2}{5}M_* r_*^2 \Omega \dot{\Omega}\ .
\label{kinetic}
\end{equation}
Here, $M_*\sim 1.4 M_{\odot}$, $r_*\sim 10$~km, and $\dot{(...)}\equiv {\rm
  d}(...)/{\rm d}t$.  Energy is lost through electromagnetic torques in the
magnetosphere, although other physical processes have at times also been
discussed (gravitational radiation, wind outflow, star-disk interaction, etc.).

To a first approximation, the stellar magnetic field may be considered as that
of a rotating magnetic dipole.  Even under such a simplification the general
description of the stellar magnetosphere is a formidable three dimensional
problem, since, in general, the magnetic and rotation axes do not coincide.
Awaiting the development of the general theory, one can still derive important
conclusions based on two idealized limiting cases which have been studied in
great extent: the case of an aligned magnetic dipole rotating in an atmosphere
with freely available electric charges (i.e. with ideal MHD conditions), and
that of a misaligned magnetic dipole rotating in vacuum.

The neutron star is not surrounded by vacuum, and one needs to take into
consideration the electric fields that develop and the electric currents that
flow in the rotating charged magnetosphere (Goldreich \& Julian~1969).  The
most recent numerical calculation of the simplest possible case, that of the
magnetosphere of an aligned rotator in force-free approximation
(Contopoulos~2005, hereafter C05), yielded the following rather general result
for the electromagnetic energy loss
\begin{equation}
L_{\rm aligned}=
\frac{4\Omega\Omega_{F}\psi_{\rm open}^2}{6c}\  
\label{C05}
\end{equation}
(the quantities $\Omega_F$ and $\psi_{\rm open}$ are defined below).  In that
picture, the magnetosphere consists of a corotating region of closed fieldlines
which extends up to a distance $r_c$ from the rotation axis, and an open
fieldline region with enclosed magnetic flux
\begin{equation}
\psi_{\rm open}\equiv \frac{1}{2\pi}\int_{\rm open}
{\bf B}\cdot {\rm d}{\bf S}
=1.23\frac{B_* r_*^3}{2r_c},
\label{psi1}
\end{equation}
where $B_*$ is the polar value of the magnetic field (Contopoulos, Kazanas \&
Fendt~1999, hereafter CKF; Gruzinov 2005; C05; Timokhin 2005; see also
Appendix~A).  The above expression is valid when $r_c\gg r_*$.  In the limit
$r_c=r_*$, straightforward calculation yields $\psi_{\rm open}=B_* r_*^2/2$.

The neutron star spins down because of the establishment of a large scale
poloidal electric current circuit flowing along open field lines, and returning
along the edge of the open field line region (see CKF for a detailed
description).  The electric current that flows between the magnetic axis
(characterized by $\psi=0$) and the edge of the open field line region
(characterized by $\psi=\psi_{\rm open}$) generates the spindown torque which
leads to eq.~\ref{C05}.  The quantity $\Omega_{F}$ in eq.~\ref{C05} is the
angular frequency of rotation of the open field lines.  It is set by the
electric potential drop that develops accross open field lines, between the
magnetic axis and the edge of the open field line region. This potential in the
magnetosphere is in general {\rm smaller} than the corresponding electric
potential drop on the surface of the star. The difference between the two
potential drops is just the particle acceleration gap potential which develops
{\em along} open magnetic field lines in the vicinity of the polar cap.
Consequently, the angular velocity of open field lines will in general be
different (smaller) than $\Omega$.  Models of particle acceleration and pair
creation of rotation-powered pulsars yield values of the gap potential $V_{\rm
  gap}$ of the order of $10^{12}$~Volts (e.g. Hibschmann \& Arons 2001).  One
can directly show (see Appendix~B) that the above together with the simplifying
assumption that $\Omega_F$ is uniform accross open field lines, yield
\begin{equation}
\Omega_F=\Omega-\Omega_{\rm death}\ ,
\label{OmegaF}
\end{equation}
where,
\begin{equation}
\Omega_{\rm death}\equiv \frac{V_{\rm gap}}{\psi_{\rm open}}c\ . \label{om_death}
\end{equation}
This describes the so-called pulsar `death', i.e. the stopping of pulsar
emission. As the neutron star slows down and $\Omega$ drops below $\Omega_{\rm
  death}$, the gap potential cannot attain the value required for particle
acceleration and consequent pulse generation, and the pulsar stops generating
radio emission.

A misaligned dipole rotating in vacuum loses energy at a rate
\begin{equation}
L_{vacuum}=\frac{B_*^2\Omega^4r_*^6}{6c^3}\sin^2\theta\ ,
\label{dipole1}
\end{equation}
where $\theta$ is the misalignment angle between the magnetic and rotation
axes.  We know that in real life the neutron star is not surrounded by vacuum,
and we may argue that, in analogy to the aligned case, the magnetosphere
consists of a corotating and an open line region. We may thus rewrite
eq.~\ref{dipole1} in a more general form that expresses the energy loss rate of
an orthogonal ($\theta=90^o$) magnetic rotator as
\begin{equation}
L_{\rm orthogonal}
\sim\frac{4\Omega^2\psi_{\rm open}^2}{6c}\ .
\label{dipole}
\end{equation}

We would like to emphasize at this point that eqs.~\ref{C05} \& \ref{dipole1}
being so similar, led most researchers to ignore the dependence on $\theta$ and
$\Omega_F$ in estimates of stellar magnetic fields, and in most studies of the
$P-\dot{P}$ diagram.  The aim of the present work is to show that the
dependence on $\theta$ and $\Omega_F$ is important and should not be ignored,
especially in old pulsars approaching their death.  As we will see, pulsar
death manifests itself in a most interesting way through its dependence on
$\Omega_F$.  Moreover, the $\theta$ dependence `softens' the distribution of
pulsars around the death line in the $P-\dot{P}$ diagram.

\section{Electromagnetic Energy Losses}

In real life, pulsars are neither aligned nor perpendicular rotators. It is
natural, therefore, to expect that the two electromagnetic energy loss terms
described in eqs.~\ref{C05} \& \ref{dipole} contribute together in the total
energy loss. A possible combination is:
\[
L\sim L_{\rm orthogonal}\sin^2\theta+
L_{\rm aligned}\cos^2\theta
\]
\[
\sim \frac{2\Omega^2 \psi_{\rm open}^2}{3c}
(\sin^2\theta+\frac{\Omega_F}{\Omega}\cos^2\theta)
\]
\begin{equation}
=\frac{B_*^2\Omega^2r_*^6}{4cr_c^2}
(\sin^2\theta+\left(1-\frac{\Omega_{\rm death}}{\Omega}\right)
\cos^2\theta)\ .
\label{Lgeneral}
\end{equation}
We stress that this combination, although inspired by physical limiting cases,
is not a rigorous derivation.  An exact expression will require the development
of a detailed non-axisymmetric MHD theory of the pulsar magnetosphere.  As we
will show, however, this rather simple general expression accounts
qualitatively for the main contributions in the electromagnetic energy loss
mechanism.

A key element in the present discussion is what determines $\psi_{\rm open}$,
or equivalently, what determines the equatorial extent $r_c$ of the closed line
region.  In the context of force-free ideal MHD, the only natural length scale
is the light cylinder distance $r_{lc}\equiv c/\Omega$.  We could have directly
assumed that $r_c$ is close to $r_{lc}$, i.e. that the corotating closed line
region extends up to the light cylinder distance (see CKF).  But this does not
answer the question how the corotating region follows the light cylinder as the
star spins down and the light cylinder moves out.  Obviously, in order for
$r_c$ to follow $r_{lc}$, the corotating closed line region must grow through
{\em north-south poloidal magnetic field reconnection around $r_c$}.  One may
consider two limiting cases:

\begin{itemize}
\item Reconnection is very efficient, and the extent of the closed 
line region follows closely the light cylinder, i.e. 
\begin{equation}
r_c\approx r_{lc}\ .
\label{rc1}
\end{equation}
\item Reconnection is very inefficient, and the closed line
region cannot grow. In that case,
\begin{equation}
r_c\approx \mbox{const.}
\label{rc2}
\end{equation}
One might argue that the supernova explosion that led to
the formation of the spinning neutron star blew the magnetic field open,
and the magnetic field remained open since then. 
\end{itemize}
We are going to consider a simple parametric model
where $r_c$ is neither equal nor proportional to $r_{lc}$,
but is equal to
\begin{equation}
r_c=r_{lc}\left(\frac{r_{lc\circ}}{r_{lc}}\right)^\alpha
=r_{lc}\left(\frac{\Omega}{\Omega_\circ}\right)^\alpha\ ,
\label{rcn}
\end{equation}
where $r_{lc\circ}$ and $\Omega_\circ$ are the values of $r_{lc}$ and $\Omega$
at pulsar birth repsectively, and $0\leq\alpha\leq 1$.  Equation~\ref{Lgeneral}
has a {\em different from dipolar} dependence on $\Omega$, i.e. a dependence
different from $\propto \Omega^4$. This accounts for braking index values
$n\neq 3$ (we introduce here the first and second order braking indices
$n\equiv \ddot{\Omega}\Omega/\dot{\Omega}^2$ and $m\equiv
\stackrel{...}{\Omega}\Omega^2/\dot{\Omega}^3$ respectively).

We have presented here a natural physical model that accounts for braking
indices $n\neq 3$, namely a corotating closed line region which does not extend
up to the light cylinder but follows the light cylinder at an increasing
distance. The dependence of energy loss on the extent of the magnetosphere and
its effect on the braking index was noted in the past (Blandford \& Romani
1988; Harding {\em et al.}~1999; Alvarez \& Carrami\~{n}ana~2004; Gruzinov
2005b; Timokhin 2005), with Gruzinov (2005b) and Timokhin (2005) introducing
the same evolutionary parametrization as in (\ref{rcn}). The physical
motivation of the extent of the closed zone varies between the authors (e.g.,
inertial stresses by the wind in Blandford \& Romani 1988 and in Harding {\em
  et al.}~1999, obliqueness effects in Gruzinov 2005b, or inability of pulsar
gaps to supply necessary poloidal currents in Timokhin 2005).  Our view,
however, that $r_c<r_{lc}$ might be due to insufficiently fast reconnection
near the Y-point in the outer magnetosphere is original and worthy of further
study.

More importantly, though, eq.~\ref{Lgeneral} 
brings a new element to the discussion:
{\em we directly introduce pulsar death and magnetic-rotation
axes misalignment in the electromagnetic pulsar spindown expression.}
Pulsar death expresses the inability of the
magnetosphere to generate the particles required in the
poloidal electric currents which generate spindown torques 
in the aligned rotator case. They do not affect the orthogonal spindown 
component though. 
Beyond the death line, the misaligned neutron star will continue to spin 
down without emitting pulsar radiation.

\begin{figure}
\includegraphics[scale=.45]{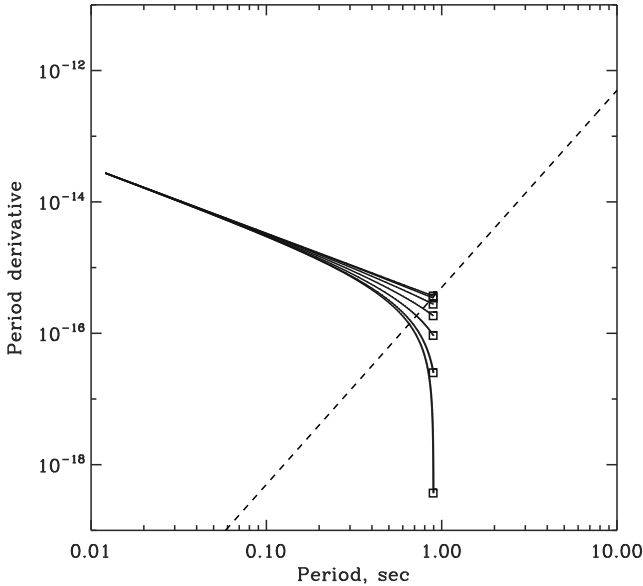}
\caption{ $P-\dot{P}$ evolutionary diagram that shows the effect of the
  misalignment angle $\theta$.  We show spindown trajectories for pulsars with
  initial period $P_\circ=10$~msec, magnetic field $B_*=10^{12}$~G, and $V_{\rm
    gap}=10^{13}$~Volt.  Initial braking index $n=3$ is assumed
  ($\alpha=0$). Trajectories from bottom to top correspond to pulsars with
  increasing magnetic inclination angle from $0^\circ$ to $90^\circ$ with
  $15^\circ$ increments. Rectangular dots indicate when pulsars reach $P_{\rm
    death}$ and turn off.  The dashed line represents the theoretical death
  line as given in eq.~\ref{logdeath} which does not take into account the
  misalignment angle dependence introduced in eq.~\ref{Pdot}. Oblique pulsars
  evolve faster through the diagram.
}
\label{fig1}
\end{figure}

\section{Period Evolution and the $P-\dot{P}$ Diagram}

In order to study how pulsars spin down, we equate the
values of $L$ in eqs.~\ref{kinetic} 
\& \ref{Lgeneral} and thus obtain
\[
\dot{P}=3.3\times 10^{-16}
\left(\frac{P}{P_\circ}\right)^{2\alpha -1}
\left(\frac{B_*}{10^{12}~\mbox{G}}\right)^2
\left(\frac{P_\circ}{\mbox{sec}}\right)^{-1}
\]
\begin{equation}
\times (\sin^2\theta+
\left(1-\frac{P}{P_{\rm death}}\right)\cos^2\theta)
\label{Pdot}
\end{equation}
where from (\ref{om_death})
\[
P_{\rm death}=8.1^{\frac{1}{2-\alpha}}\ \mbox{sec}
\]
\begin{equation}
\times \left(\frac{B_*}{10^{12}\mbox{G}}\right)^{\frac{1}{2-\alpha}}
\left(\frac{V_{\rm gap}}{10^{12}\mbox{Volt}}\right)^{-\frac{1}{2-\alpha}}
\left(\frac{P_\circ}{\mbox{sec}}\right)^{-\frac{\alpha}{2-\alpha}}\ .
\end{equation}
$P_\circ$ is the initial period at pulsar birth.  Note that we have introduced
an angular dependence in the definition of the death line $\dot{P}(P_{\rm
  death})$.  As we will see below, this has interesting observational
consequences. Note also that, in the limit $\theta\sim 90^\circ$ and
$\alpha=0$, we obtain the following simple expression
\begin{equation}
\dot{P}_{\rm death}=5\times 10^{-18}
\left(\frac{P_{\rm death}}{P_\circ}\right)^{3}
\left(\frac{V_{\rm gap}}{10^{12}\mbox{Volt}}\right)^{2}
\left(\frac{P_\circ}{\mbox{sec}}\right)^{3}\ ,
\end{equation}
which can also be written as
\begin{equation}
\log\dot{P}_{\rm death}=3\log P_{\rm death}-17.3+2(\log V_{\rm gap}-12)
\label{logdeath}
\end{equation}
(compare with Zhang, Harding \& Muslimov~2000).
When $P\ll P_{\rm death}$, 
\begin{eqnarray}
B_* & \propto & \dot{P}^{\frac{1}{2}}P^{\frac{1}{2}-\alpha}\\
n & = & 3-2\alpha\\
m & = & n(n-1)\ ,
\end{eqnarray}
and pulsars evolve in the logarithmic $P-\dot{P}$ diagram along lines of
constant magnetic field $\dot{P}\propto P^{2\alpha -1}$.  As $P$ approaches
$P_{\rm death}$, however, the evolution curves down away from the above
straight lines and ends at a point $P=P_{\rm death}$ and
$\dot{P}=\dot{P}(P_{\rm death})$ (see figure~\ref{fig1}).  This evolution is
similar to what one would expect if we had assumed magnetic field decay in the
standard dipole spindown model (e.g. Gonthier~{\em et al.}~2002 and 2004,
hereafter G02 and G04 respectively).  In figure~\ref{fig1} we show only
$\alpha=0$, or evolution with initial braking index $n=3$. For $n<3$ the
straight portion of the lines would rotate counterclockwise, and would become
horizontal for $n=2$ ($\alpha=0.5$). Such pulsars would also evolve faster.

As a test of the validity of our results, we tried to reproduce the observed
distribution of pulsars in the $P-\dot{P}$ diagram with the minimum number of
assumptions possible, through a simple Monte Carlo numerical experiment that
follows G02 and G04.  We integrated eq.~\ref{Pdot} for random times $t$ within
a $10^9$~year time interval (this mimics random pulsar birth in the
galaxy). The result of the integration is recorded only if the star remains
active as a pulsar, i.e.  only if $P(t)\leq P_{\rm death}$.  To make our
results more realistic, we assumed a multipeaked lognormal distribution of
polar magnetic fields around $B_*\sim 10^{12}\mbox{G}$ (eq.~1 in G02):
$\rho_B=\Sigma_{i=1}^2 A_i e^{-(\log B-\log B_i)^2/\sigma^2_i}$.  We used two
terms with $A_1=20.0$, $A_2=2.0$, $\log B_1=12.65$, $\log B_2=13.0$,
$\sigma_1=0.6$, and $\sigma_2=0.4$.

We assumed a uniform random distribution of initial pulse periods from 10~msec
to 0.2~sec.  We did account for the observational selection effect due to
finite instrument detectability by implementing the radio pulsar luminosity
model of Narayan \& Ostriker~1990 (eqs.~23, 24 in G02), and a lower detectable
pulsar luminosity limit of $12.5$~mJy$\cdot$kpc$^2$.  Finally, we assumed that
there is no magnetic field or misalignment angle evolution with time.  In an
effort to simplify the approach we did not consider the multitude of selection
effects and individual survey detectability limits. We also did not calculate
the distance to individual pulsars after tracking them in the galactic
potential as in G02 and G04, relying on the luminosity distribution
function. We believe, however, that the gross effects of the new spindown law
(\ref{Pdot}) can be understood even with our procedure.

\begin{figure}

\includegraphics[scale=.45]{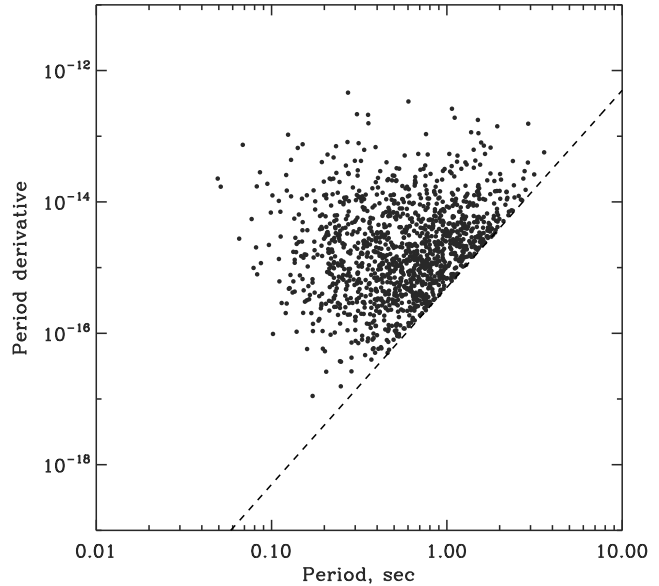}
\caption{Monte Carlo experiment using the standard dipole model ($\alpha=0$, or
  equivalently $n=3$, and no angle dependence).  Pulsars are injected with a
  lognormal distribution of polar magnetic field values around $B_*=10^{12}$~G
  (as in G02).  $V_{\rm gap}=10^{13}$~Volt.  $P_\circ$, the initial period at
  pulsar birth, is uniformly randomly distributed between 10~msec and 0.2~sec.
  We have also introduced an observational selection effect due to finite
  instrument detectability limit (see text).  The dashed line represents the
  theoretical death line as given in eq.~\ref{logdeath}.  The strong
  concentration of pulsars near the death line is not observed in the real
  distribution of pulsars.  }
\label{fig2}
\end{figure}

\begin{figure}
\includegraphics[scale=.45]{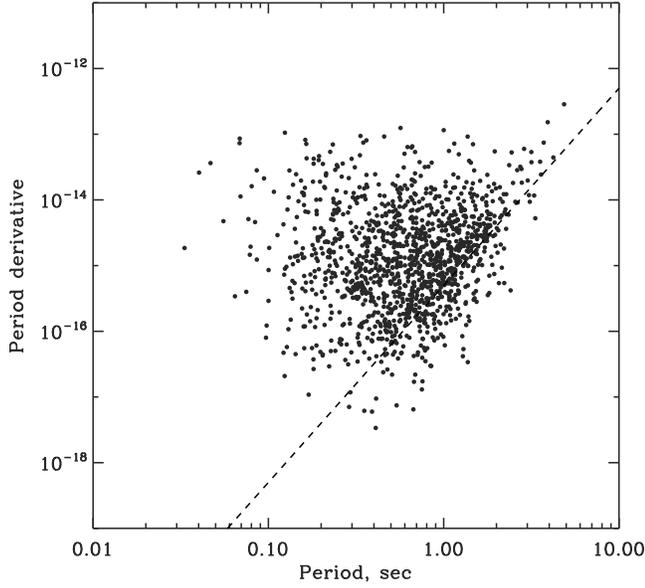}
\caption{Same as fig.~\ref{fig2}, only with the misalignment
random angle dependence obtained in eq.~\ref{Lgeneral}.
The introduction of the angle dependence results in
the softening of the death line.
}
\label{fig3}
\end{figure}

\begin{figure}
\includegraphics[scale=.45]{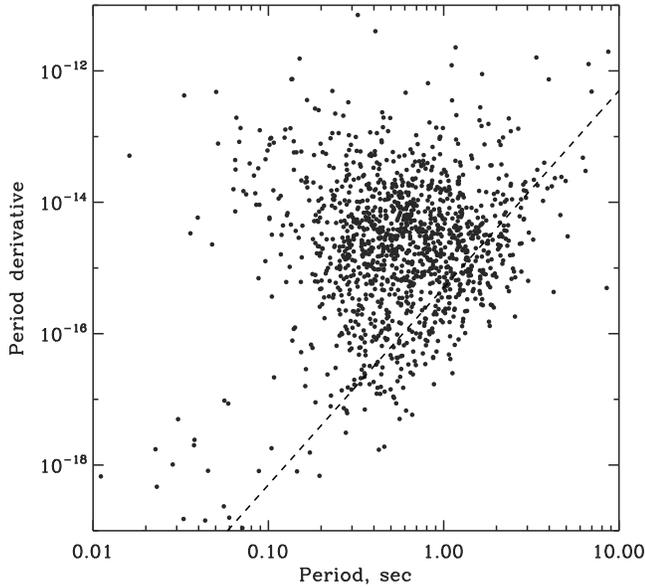}
\caption{The present day observed radio pulsar distribution. Data from ATNF
  pulsar catalog.  }\label{fig4}
\end{figure}

Our results are shown in figs.~2,3 \& 5.  In figure~\ref{fig2} we show the
result of our simulation for the standard dipolar spindown ($\alpha=0$, or
equivalently $n=3$, and no dependence on misalignment angle $\theta$, i.e.,
$\theta=90^\circ$ for all pulsars). One can directly see a strong concentration
of pulsars near the death line, although the concentration is decreased by the
luminosity selection effect, which favors younger brighter pulsars and fills up
the upper left corner of the diagram.  This concentration is understood since
pulsars evolve slower and slower as they grow older. This is, however, not what
is observed (figure~4).  The need to make the death line boundary softer led
some researchers to introduce a "death valley" (G02) or field decay (G02, G04)
in their simulations. In figure~\ref{fig3} we show that introducing the
$\theta$ dependence that we have obtained in eq.~\ref{Lgeneral} with random
misalignment angles, solves this problem naturally without magnetic field
decay.  As we saw in figure~\ref{fig1} the angle-dependent death line helps in
smoothing the pileup.
In figure~\ref{fig4}, we show the present day observed radio pulsar
distribution from the ATNF pulsar catalog\footnote{Available at
  http://www.atnf.csiro.au/research/pulsar/psrcat/} for comparison with our
results.  The similarity between figures~\ref{fig3} \& \ref{fig4} is quite good
considering the simplifications of our model. We aim to reproduce the gross
features of the observed $P-\dot{P}$ diagram for non-millisecond pulsars -- its
triangular shape and centroid location, while keeping the number of initial
magnetic field components to a minimum. In the standard spindown picture it is
very difficult to have pulsars in the lower corner of the $P-\dot{P}$ triangle
without polluting the low $P$ -- low $\dot P$ region of the diagram with
low-field pulsars, which are not observed. In our spindown model the lower
corner is filled with pulsars dropping from the higher $\dot P$ (higher $B_*$)
region, rather than the ones moving on straight tracks as in the standard
case. This contributes to the triangular shape of the diagram.

\begin{figure}
\includegraphics[scale=.45]{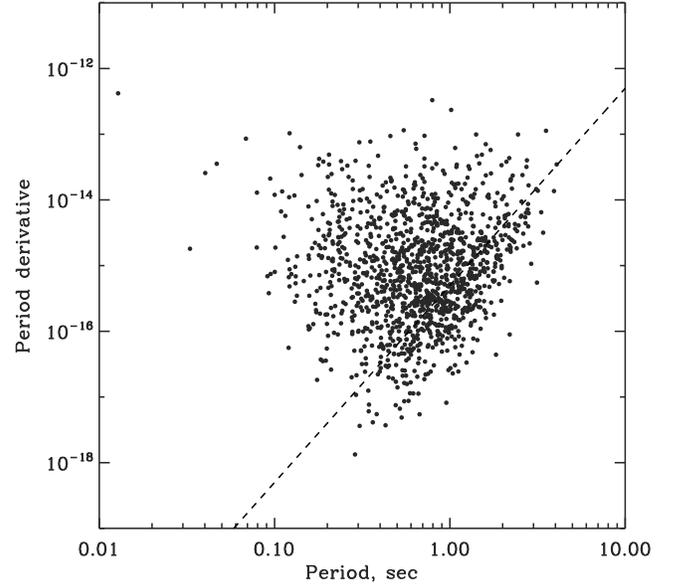}
\caption{Same as figure~2, but with misalignement angles constrained to be less than $45^\circ$.}
\label{fig41}
\end{figure}

We obtain a better overall fit if we constrain the misalignment angles to be
smaller than $45^\circ$ (figure \ref{fig41}). This reduces the horizontal width
of the triangle, which means that more higher-field pulsars are moving on
vertical tracks in the diagram. Also, the ridge of overdensity near the
theoretical death line seen in figure~\ref{fig3} is smoothed out in
figure~\ref{fig41}.

Similar shapes of the diagram are found for $\alpha>0$.  This explains
individual observed pulsars with $n<3$.  However, our fit deteriorates for
values of $\alpha$ larger than 0.25. We thus conclude that, although individual
pulsars may well have $\alpha>0$, as a whole, $\alpha \sim 0$ is a good
approximation for the observed pulsar population.

Furthermore, when $\alpha=0$, from eqs.~\ref{kinetic} \& \ref{Lgeneral} we
derive
\begin{equation}
n=3+\frac{2}{\Omega/(\Omega_{\rm death}\cos^2\theta)-1}\ .
\end{equation}
Around the pulsar death line
$\Omega\approx \Omega_{\rm death}$, this equation gives
\begin{equation}
n\approx 3+\frac{2 \cos^2\theta}{1-\cos^2\theta}\ ,
\end{equation}
which, for aligned pulsars with $\cos\theta\approx 1$ may be
much greater than 3. Such high braking index values may be observable (Johnston \& Galloway 1999).

\begin{figure}
\includegraphics[scale=.45]{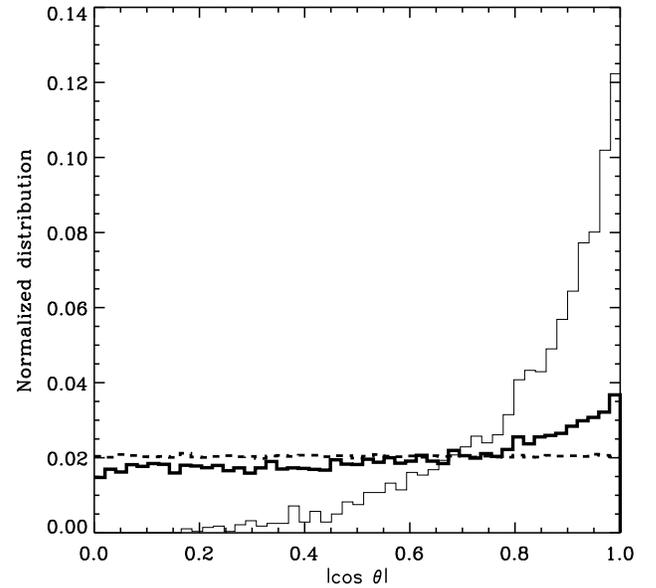}
\caption{The distribution of $|\cos(\theta)|$ in our Monte Carlo experiment in
  figure~3. The dashed line represents the original distribution injected with
  uniform random misalignment angle.  The thick solid line represents the
  distribution of misalignment angles in the simulated pulsar distribution as a
  whole. The thin solid line represents the distribution of misalignment angles
  for the pulsars observed lying below the theoretical death line given by
  eq.~\ref{logdeath}. An interesting observational prediction of the present
  work is that pulsars lying near or below the theoretical death line should
  have preferentially smaller misalignment angles.  }
\label{fig5}
\end{figure}

In figure~\ref{fig5} we plot the distribution of $\cos\theta$ in our Monte
Carlo simulation that was presented in figure~\ref{fig3}. In this simulation
pulsars are injected (born) with random magnetic axis orientation with respect
to the rotation axis as shown by the dashed histogram in figure ~\ref{fig5}.
The pulsar population at the end of the simulation, however, shows a different
distribution of misalignment angles as shown by the thick solid line.  The
selection bias towards smaller misalignment angles may be attributed to the
fact that orthogonal rotators spin down faster (on the average), and therefore
are more likely to evolve beyond the death line within the observation time
interval.  Finally, the thin solid line represents the distribution of
misalignment angles in the pulsars observed lying below the theoretical death
line given by eq.~\ref{logdeath} (which does not take into account the angle
dependence introduced in eq.~\ref{Pdot}). One sees clearly that {\em pulsars
  observed near or below the death line should have preferentially smaller
  misalignment angles $\theta$ than the rest of the pulsar population}. This
is, we believe, an important observational prediction of the present work.


\begin{figure}
\includegraphics[scale=.45]{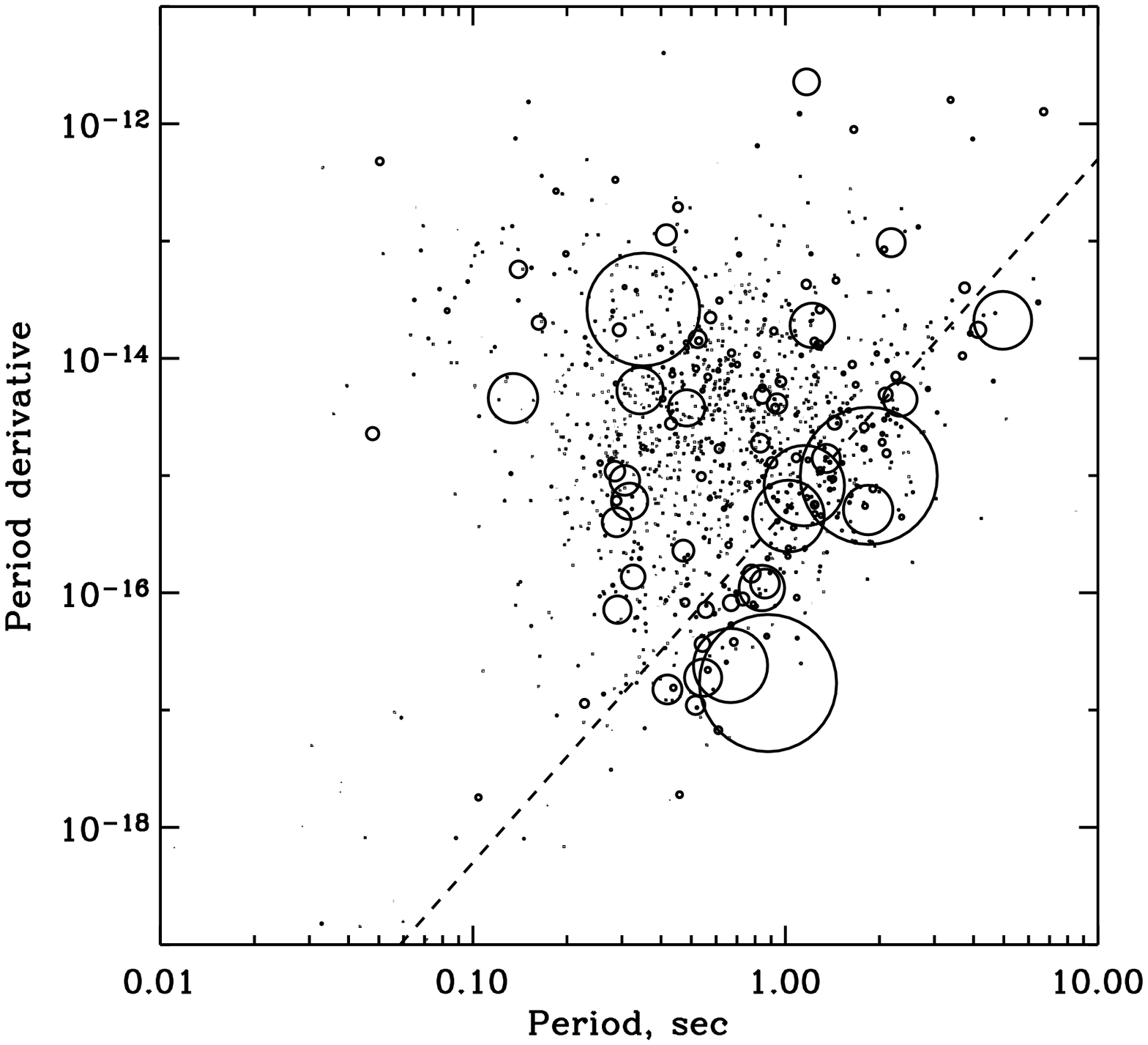}
\caption{Distribution of alignment measure in the observed pulsar
  population. Each pulsar is plotted as a circle with radius proportional to
  the quantity $W P^{-1/2}$ using the pulse width $W$ at 50\% of the peak.
  Data from ATNF pulsar catalog.  }
\label{fig51}
\end{figure}

\section{Discussion}

Awaiting the development of a detailed three dimensional MHD theory for the
rotating neutron star magnetosphere, we can describe a few general
characteristics of its expected structure.

We argue that $\Omega_F$ characterizes the reduced magnetospheric electric
potential drop between the magnetic axis and the edge of the open field line
region.  Poloidal electric currents will be generated as long as electric
charges can be produced in the magnetosphere.  These charges are produced in
the magnetospheric polar gaps.  In the axisymmetric case, $\Omega_F$ can also
be thought as the angular velocity of rotation of open magnetic field lines,
which is in general smaller than $\Omega$.  The electric current that flows
between the magnetic axis and the edge of the open field line region generates
the spindown torque. As shown in C05, this electric current and spindown torque
are both proportional to $\Omega_F$.

We argue that this picture describes also the situation when $\theta\neq 0$.
As the neutron star rotates, the magnetic axis moves around the axis of
rotation. In the rotating frame of the star, however, open field lines rotate
at their own rate $\Omega-\Omega_F$ opposite to the direction of stellar
rotation around the magnetic axis.  Observational evidence for this effect can
be found in the well known sub-pulse drift phenomenon (e.g. Beskin 1997; Rankin
\& Wright 2003).  In the limit $\Omega_F=\Omega$, it is as if open magnetic
field lines are anchored on the stellar surface.  In the opposite limit
$\Omega_F=0$, open field lines rotate around both the magnetic and rotation
axes returning to the same position every rotation of the star when viewed by
an inertial observer. For finite $\Omega_F $ a phase shift about the magnetic
axis will be accumulated after every turn of the star.  A simple analogy to the
above picture might be that of a lawn watering system consisting of a hose
rotating at angular velocity $\Omega$, and a sprinkler at its end rotating with
angular velocity $\Omega_F-\Omega$ in the rotating frame of the watering hose.

Note that the orthogonal component does not require
the establishment of a poloidal electric current
circuit in order for it to slow down the stellar rotation.
The orthogonal component emits spiral electromagnetic waves 
(in general Alfven waves) which travel out to infinity
through the rotating or non-rotating open field lines.

The main conclusions of the present work are:
\begin{enumerate}
\item The electromagnetic energy loss of a pulsar depends not only
on the surface magnetic field $B_*$ and on the rotational
angular velocity $\Omega$, but also on the misalignment angle
$\theta$ and on the angular velocity of the open magnetic field
lines $\Omega_F$ (which is in general smaller than $\Omega$).
\item The approach to pulsar death modifies the rate of energy loss. The death occurs
 when the pulsar slows down sufficiently so that the angular velocity of the open fieldlines
$\Omega_F=0$.
\item The energy loss close to the pulsar death is smaller
than what is given by the standard dipolar spindown formula.
This effect gives a good fit between the theoretical and
observed distributions of pulsars near the death line,
without invoking a magnetic field decay.
\item Our model may also account for individual pulsars
spinning down with braking index $n<3$. However, $n\sim 3$ remains
a good approximation for the pulsar population as a whole.
\item Pulsars near the death line have braking index values $n> 3$. 
Such high braking index values may be observable.
\item Pulsars near the death line may have preferentially smaller inclination angles. 
\end{enumerate}

A preliminary look at the ATNF pulsar catalog data suggests that the last point
may have some observational support. One possible measure of the inclination
angle of a pulsar is its fractional pulse width, or the ratio of the width of
the pulse to the period of the pulsar. If the radio beam size is independent of
inclination, then we would expect pulsars with smaller inclinations to be seen
for a larger fraction of the period than pulsars with large inclinations. In
order to test this hypothesis we took the available data for the pulse width at
50$\%$ of the pulse peak from the ATNF catalog (1375 pulsars with $W_{50\%}
\neq 0$).  In order to be able to compare the fractional pulse width for
pulsars of different periods, we have to correct for the intrinsic size of the
pulsar beam.  If the beam roughly follows the angular size of the open
fieldlines, then the beam size falls as $P^{-1/2}$ with increasing period (we
are assuming that the last closed field line extends out to the light
cylinder). Therefore, the quantity that should relate to the degree of
alignment is $F_{\rm align} \equiv (W\times P^{1/2})/P=W P^{-1/2}$, where $W$
is the measured pulse width. In figure~\ref{fig51} we plot the observed pulsars
as circles with the radius of the circle linearly proportional to $F_{\rm
  align}$.  A visual inspection of the plot shows that there is an excess of
larger pulse fractions for older pulsars, and in particular for pulsars near
the right edge of the $P-\dot{P}$ triangle. This region is near the pulsar
death line (dashed line), and therefore, this feature is particularly
interesting. Obviously, a much better analysis needs to be performed, but these
results are quite encouraging.  An overabandunce of pulsars with large pulse
fractions near the death line in a smaller pulsar sample had also been
interpreted by Lyne and Manchester (1988) as an indication of alignment.

In an effort to simplify the fits we have assumed in this paper that the
misalignment angle stays the same during the evolution, and there is no field
decay. The reality may of course be more complicated, and there could be some
amount of field decay, and potential alignment (and consequent precession)
during the lifetime of a pulsar, due to electromagnetic torques on the
star. These effects would introduce extra degrees of freedom to fitting the
pulsar distribution. We hope, however, that the physically motivated spindown
law of the type introduced in this work would find its way into detailed
population synthesis models.

\acknowledgements{ We would like to thank J. Arons, D. Backer, R. Blandford,
  G. Contopoulos, J. Granot, D. Psaltis, R. Romani, J. Seiradakis, and
  A. Timokhin for useful conversations and suggestions during the preparation
  of this work.  A.S. acknowledges support provided by NASA through Chandra
  Fellowship grant PF2-30025 awarded by the Chandra X-Ray Center, which is
  operated by the Smithsonian Astrophysical Observatory for NASA under contract
  NAS8-39073.  }

\begin{appendix}
\section{Appendix A: Calculation of Open Magnetic Flux}

We present here expressions for $\psi_{\rm open}(r_c)$
obtained numerically for various axisymmetric
magnetospheric configurations using the numerical code developed
in C05: 
\begin{itemize}
\item  Standard magnetostatic dipole:
\begin{equation}
\psi_{\rm open}=\frac{B_* r_*^3}{2r_c}\ .
\end{equation}
Note that in this limit, $\psi_{\rm open}(r_c)$ is defined
as the amount of flux that crosses the equator beyond equatorial
distance $r_c$. All field lines cross the equator perpendicularly.
\item Magnetostatic dipole
with open field lines beyond $r_c$:
\begin{equation}
\psi_{\rm open}=1.20\frac{B_* r_*^3}{2r_c}\ .
\label{A2}
\end{equation}
Note that this implies the existence of an
equatorial azimuthal current sheet beyond $r_c$. The magnetosphere
is otherwise stress-free. The last open field lines crosses
the equator at an angle of about $70^o$ towards the axis
(`Y' point).
\item Rotating (relativistic) dipole
with open field lines beyond $r_c$, and smooth crossing of 
the light cylinder:
\begin{equation}
\psi_{\rm open}=1.23\frac{B_* r_*^3}{2r_c}\ .
\end{equation}
This result is valid for 
$r_*\ll r_c\leq r_{lc}$. 
The solution requires the presence of a) a poloidal
electric current along open field lines, b) a poloidal current sheet
along the separatrix between open and closed field lines, and c)
an equatorial azimuthal current sheet beyond $r_c$.
The last open field line crosses the equator at an angle
of about $70^o$ towards the axis.
Note the interesting similarity with the latter non-relativistic
case (eq.~\ref{A2}).
\end{itemize}


\section{Appendix B: Calculation of $\Omega_F$}

We derive here an approximate expression for $\Omega_F$, assuming
$\Omega_F=\mbox{const}$. We perform the calculation in 
the aligned (axisymmetric) case $\theta=0$.
It has been shown in C05 that $\Omega_F$ is related to the 
magnetospheric potential drop $V_F$ between
the magnetic axis (characterized by $\psi=0$) and the edge of the 
open field line region
(characterized by $\psi=\psi_{\rm open}$), namely
\[
V_F=\int_{\psi=0}^{\psi_{\rm open}}{\bf E}\cdot{{\rm d}{\bf s}}
=\int_{\psi=0}^{\psi_{\rm open}}
\frac{s\Omega_F}{c}B_p{\rm d}s
=\frac{1}{c}\int_{\psi=0}^{\psi_{\rm open}} \Omega_F {\rm d}\psi
\]
\begin{equation}
=\frac{\Omega_F \psi_{\rm open}}{c}\ ,
\label{Appendix1}
\end{equation}
where $s$ is the distance from the rotation axis; 
${\bf E}$ is the electric field; $B_p$ is the poloidal
(meridional) component of the magnetic field. $V_F$ is in general
{\em different} from the corresponding
stellar potential drop, namely
\begin{equation}
V_* =\frac{1}{c}\int_{\psi=0}^{\psi_{\rm open}} \Omega {\rm d}\psi
=\frac{\Omega \psi_{\rm open}}{c}\ .
\label{Vstar}
\end{equation}
The difference 
\begin{equation}
V_{\rm gap}\equiv V_*-V_F=\frac{(\Omega-\Omega_F)\psi_{\rm open}}{c}
\label{V1}
\end{equation}
is just the particle acceleration gap potential which devolops
{\em along the magnetic field} near 
the footpoint of open field lines (e.g. Beskin~1997).
Therefore,
\begin{equation}
\Omega_F=\Omega-\Omega_{\rm death}\ ,
\end{equation}
where
\begin{equation}
\Omega_{\rm death}\equiv \frac{V_{\rm gap}}{\psi_{\rm open}}c\ .
\end{equation}
This expression is valid as long as $\Omega\geq  \Omega_{\rm death}$.
When $\Omega< \Omega_{\rm death}$, $\Omega_F=0$.
\end{appendix}

\end{document}